# Electrostatic force microscopy and potentiometry of realistic nanostructured systems

*M. Lucchesi*[(1,2)], *G. Privitera*[(1)], *M. Labardi*[(2)]\*, *D. Prevosto*[(2)], *S. Capaccioli*[(1,2)], *P. Pingue*[(3)]

[1] Dipartimento di Fisica "Enrico Fermi", Università di Pisa, Largo Pontecorvo 3,56127 Pisa, Italy

[2] INFM-CNR polyLab, Largo Pontecorvo 3,56127 Pisa, Italy

[3] INFM-CNR NEST, Scuola Normale Superiore, Piazza S. Silvestro 12, 56127 Pisa, Italy

\* labardi@df.unipi.it; Tel. +39-050-2214322: Fax +39-050-2214333

Running title: Electrostatic force microscopy of nanostructures.

We investigate the dependency of electrostatic interaction forces on applied potentials in Electrostatic Force Microscopy (EFM) as well as in related local potentiometry techniques like Kelvin Probe Microscopy (KPM). The approximated expression of electrostatic interaction between two conductors, usually employed in EFM and KPM, may loose its validity when probe-sample distance is not very small, as often realized when realistic nanostructured systems with complex topography are investigated. In such conditions, electrostatic interaction does not depend solely on the potential difference between probe and sample, but instead it may depend on the bias applied to each conductor. For instance, electrostatic force can change from repulsive to attractive for certain ranges of applied potentials and probe-sample distances, and this fact cannot be accounted for by approximated models. We propose a general capacitance model, even applicable to more than two conductors, considering values of potentials applied to each of the conductors to determine the resulting forces and force gradients, being able to account for the above phenomenon as well as to describe interactions at larger distances. Results from numerical simulations and experiments on metal stripe electrodes and semiconductor nanowires supporting such scenario in typical regimes of EFM investigations are presented, evidencing the importance of a more rigorous modelling for EFM data interpretation. Furthermore, physical meaning of Kelvin potential as used in KPM applications can also be clarified by means of the reported formalism.

In electrostatic force microscopy (EFM) [1] and related local potentiometry [1,2] and Kelvin probe microscopy (KPM) [3] techniques, local electrical properties of surfaces are investigated by detecting electrostatic interactions between the sharp conducting probe of an atomic force microscope (AFM) and the sample surface, when both of them are properly biased. In EFM, electrostatic force gradients $dF_{el}/dz$ affect the behavior of the probe, vibrating along the direction $z$ normal to the surface through a flexible cantilever of spring constant $k$, dithered at frequency $\nu$ by means of a piezoelectric actuator. For small oscillation amplitude $A$ compared to the spatial range of the interaction force gradient, the latter can be linearized to yield an effective spring constant $k_{eff} = k - dF_{el}/dz$, as customarily done in non-contact atomic force microscopy [4]. This change of spring constant translates into a shift of the free resonant frequency $\nu_0$ of the cantilever of an amount $\Delta\nu >> - (\nu_0/2k)dF_{el}/dz$. Therefore, in EFM, $\Delta\nu$ provides a measurement of the electrical interaction between the probe and the local portion of surface in front of it, when both are polarized by a DC bias.

In KPM instead, the applied bias is comprised of a DC component $\varphi_1$ and an AC component $\varphi_{AC} \cos (2\pi\nu_{mod} t)$, where $\nu_{mod}$ is usually chosen equal to $\nu_0$ to enhance sensitivity by exploiting cantilever resonance. The so called "Kelvin" potential $\varphi_1$ is adjusted by a feedback control that acts in order to nullify the electrostatic force component $F_{mod}$ at frequency $\nu_{mod}$ [2,3]. In this way, it turns out that KPM is sensitive to $F_{el}$ instead than to its gradient [3].

An expression of electrostatic force in terms of applied potentials, as well as of geometric and electrical characteristics of probe and sample, is necessary to correlate results of EFM and KPM measurements to the actual structure of the sample. An approximate expression for such interaction, valid in the limit case of parallel plate capacitor and used in most of the literature about EFM is:

$$F_{el} = \frac{1}{2}\frac{dC}{dz}\Delta\varphi^2 \quad (1)$$

with $C$ the capacitance of the probe-sample system, $\Delta\varphi$ their potential difference, and $z$ their relative distance. As well known, Eq. (1) holds only for very small probe-sample distance compared to the characteristic size of both probe and sample; however, it does not provide a general description of electrostatic interactions of a two-conductor system. Strictly speaking, Eq. (1) was initially introduced in EFM as an approximation to describe the electrical behavior of flat samples. Later on, investigation of nanostructures with more complex topography and larger probe scanning distance from the surface have often led to inappropriate use of Eq. (1), whereas a more rigorous application of the principles of electrostatics would have been required.

To evidence by a simple argument the application limits of Eq. (1), as well as to clarify the basic concepts of electrostatics that will be useful for our work, let us consider two conducting spheres of equal radius $R$ and charged by equal positive charges $Q$ when at very large distance from each other (Fig. 1a). For symmetry reasons, potential difference between them must be zero for any distance $h$. According to Eq. (1), $F_{el}$ should be null; on the contrary, we know from Coulomb's law that the two spheres repel each other (force $F_a$ in Fig. 1a).

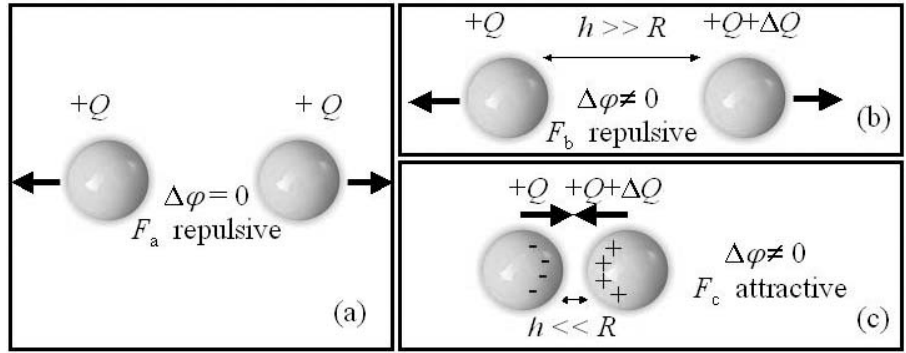

**Figure 1**: Scheme of the ideal experiment proposed to assess the limits of validity of Eq. (1). Symmetric system (a), asymmetric system at large (b) and small distance (c).

Let us now add a positive charge $\Delta Q$ to one sphere (Fig. 1b and 1c) so that such symmetry is broken. Now, at small distance $h \ll R$, attractive force $F_c$ due to mutual induction of charges of opposite sign will prevail (Fig. 1c), as correctly described by Eq. (1), whereas that is not the case at large distance $h \gg R$ (Fig. 1b), where force $F_b$ is again repulsive as in the symmetric case.

A similar behavior occurs when the two spheres are connected to voltage generators. In particular, by applying potentials $\varphi_1$ and $\varphi_2$ in such a way that a fixed $\varphi_2 - \varphi_1 = \Delta\varphi$ is maintained, attractive force is always exerted when potentials are of opposite sign, while it can be seen that force can be either repulsive or attractive for certain values of potentials when they are of the same sign, again because of induction effect dominance at small distance.

Dependency of EFM signal from the choice of potentials applied to both probe and sample has been recently observed and investigated by Qi *et al*. [5]. To describe such experimental findings, and more generally to take into account the effect of static charges, *ad hoc* Coulombic terms of the form $Q_{tip}E_{sample}$ are often added to Eq. (1) [5,6], although somehow arbitrarily. More generally, electrostatic forces in EFM may depend on potentials applied to both probe and sample, and not simply on their difference. In particular, the choice of connecting sample [1] or probe [7] to ground, or biasing both of them by independent generators referred to ground [5], or by a floating generator [8], can affect measured forces as well as their gradients.

From all the above arguments, it is evident that a more general framework is needed to describe electrostatic interactions, at variance of potentials applied to both probe and sample, in EFM and related techniques. In this paper, we present a general method to model electrostatic interactions to improve interpretation of EFM as well as of local potentiometry measurements. We will support our results by numerical simulations as well as by EFM test measurements on flat metal electrodes as well as on nanostructures like semiconductor nanowires. We finally discuss implications for KPM and local probe potentiometry.

Let us start with the description of our method. For a system of $N$ conductors, after Ref. [9] the linear relation between charge and electric potential on each conductor is:

$$Q_i = \sum_{j=1}^{N} C_{ij} \varphi_j \qquad (2)$$

where $\varphi_j$ is the potential of the $j$-th conductor, the diagonal elements $C_{ii} > 0$ are the capacity coefficients and the off-diagonal elements $C_{ij} = C_{ji} < 0$ (with $i \neq j$) are the electrostatic induction coefficients. To work out $C_{ij}$s, one should consider the inverse matrix $S_{ij}$ relating potentials to charges as:

$$\varphi_i = \sum_{j=1}^{N} S_{ij} Q_j \, . \qquad (3)$$

Here, by assuming a charge $Q_i$ on the $i$-th conductor and zero charge on all others, contributions to the potential on each conductor can be calculated by analytical or numerical methods, and consequently the $S_{ij}$ coefficients can be obtained. $C_{ij}$s are then carried out by inverting the $S_{ij}$ matrix.[1]

The force $F_z$ acting along $z$ on one of the conductors – typically the probe, acting as the force sensor, labelled here by the index "1" – is calculated as $-dU_{tot}/dz$, where $U_{tot}$ is the total energy of the system. From general grounds, $U_{tot}$ corresponds to the electrostatic energy

$$U_{el} = \frac{1}{2}\sum_{i,j=1}^{N} C_{ij}\varphi_i \varphi_j \qquad (4)$$

in case charged conductors are isolated. When potentials are kept constant by means of external generators, their work must be taken into account, leading instead to $U_{tot} = -U_{el}$ [9]. The force gradient $F'_z$ – where prime superscript indicates $z$-derivative – is obtained by further $z$-derivation of $F_z$. All our considerations in this paper are made by assuming the value of ground potential equal to the one at infinity, $\varphi_\infty = 0$. Obviously, the physics involved, and particularly the values of electrostatic forces as well as of their gradients, shall not depend on the choice of $\varphi_\infty$.

Let us now describe the simplest EFM configuration as a two-conductor system, where the probe is biased by an external generator at potential $\varphi_1$ and the conducting surface by a second generator at potential $\varphi_2$. All cases previously mentioned, except the one with floating generator between probe and sample, that however is rarely reported, can be described in such way. The force gradient experienced by the probe along $z$ results in this case:

---

[1] A similar model of electrostatic interactions between $N$ conductors to describe KPM was introduced by Jacobs et al. [10], assuming that potentials and charges were related by capacitance coefficients $C_{ij}$ as $Q_i = \sum_{j=1}^{N} C_{ij} (\varphi_i - \varphi_j)$. However, such model does not succeed to overcome the difficulties related to Eq. (1). For instance, in the case of a two-conductor system as the simple one of Fig. 1a, assuming $\varphi_1 = \varphi_2$ leads to $Q_1 = Q_2 = 0$.

$$F_z' = \frac{1}{2}C_{11}''\varphi_2^2\left[\left(\frac{\varphi_1}{\varphi_2}\right)^2 + 2\frac{C_{12}''}{C_{11}''}\frac{\varphi_1}{\varphi_2} + \frac{C_{22}''}{C_{11}''}\right], \tag{5}$$

showing quadratic dependency on the ratio $\varphi_1/\varphi_2$. By imposing $F'_z = 0$ we get the solutions:

$$\left(\frac{\varphi_1}{\varphi_2}\right)^{**} = -\frac{C_{12}''}{C_{11}''}\left(1 \pm \sqrt{\Delta''}\right), \tag{6}$$

where the double star indicates solutions that null force gradient, with

$$\Delta'' = 1 - \frac{C_{11}''C_{22}''}{C_{12}''^2}. \tag{7}$$

By means of such description we observe that, when $\Delta'' > 0$, force gradient at a given probe-sample distance can be even inverted within a certain range of potentials. In practical terms this could mean that the sign of electrostatic force gradients, and consequently the value of measured electric polarizability, could be misinterpreted in EFM, depending on the sample and probe potentials employed.

As a first predictive application of the model, let us sketch our simplified EFM system as two conducting spheres, with radii $R_1 \leq R_2$, since this case is easier to handle analytically, at least for large distance, as well as numerically for smaller distance. In this sketch, the smaller sphere could represents the probe apex and the bigger one the sample features on the surface, although it is well known that contributions from larger probe and sample portions are relevant for the overall interaction [11]. However, to understand the general properties related to our modellization, let us evaluate capacitance coefficients and their derivatives in this simple case, and analyze their behavior for varying distance and size of probe and sample. On such basis, some anticipation of the related effects on force gradient measurements in EFM could be given.

For distance $r \gg R_c = \sqrt{R_1 R_2}$,[2] one gets:

$$\begin{pmatrix} Q_1 \\ Q_2 \end{pmatrix} = C_1 \begin{pmatrix} \left(1+\dfrac{1}{\rho^2}\right) & -\dfrac{\sqrt{K_r}}{\rho} \\ -\dfrac{\sqrt{K_r}}{\rho} & K_r\left(1+\dfrac{1}{\rho^2}\right) \end{pmatrix} \begin{pmatrix} \varphi_1 \\ \varphi_2 \end{pmatrix}. \qquad (8)$$

by defining the size ratio $K_r = R_2/R_1 \geq 1$ and the normalized distance $\rho = r/R_c \geq 0$, and where $C_i = 4\pi\varepsilon_0 R_i$ are the capacitances of the isolated spheres.

For $\rho$ not much bigger than unity, the capacitance matrix cannot be expressed in analytical form anymore. However, in the case of spherical shape, it can be easily calculated numerically through the recursive image charge method [9]. Simulations by such method show that capacitance coefficients, as well as their derivatives, are always monotonically decreasing functions of distance. In particular, for smaller $\rho$, we find that all capacitance coefficient derivatives tend to assume the same value and trend. Therefore, from Eq. (5), the force gradient can always be expressed for small distance as $F'(\rho) \sim \frac{1}{2} C''_{11}(\rho)(\varphi_1 - \varphi_2)^2$. In such case, interaction between induced charges is dominant. As we will see briefly on, for small $\rho$, also force turns out to be always attractive, in agreement with Eq. (1).

As an example, we show in Fig. 2 numerically calculated values of $C''_{ij}$ in the case of two spheres. The size of the first sphere was chosen as $R_1 = 20$ nm to represent a typical probe apex radius of curvature, while the second one is $R_2 = 100$ nm, to represent a realistic nanostructure with size bigger than the probe apex. We notice how both $C''_{11}$ and $-C''_{12}$ tend to zero for large distance, and assume the same value while tending to infinity for small distance. The latter property can be demonstrated analytically for the capacitance coefficient derivatives, as follows. Let us first consider the $S_{ij}$ matrix of

---

[2] $r$ represents the distance between the surfaces of conductors, i.e. in case of two spheres the distance between centers $r_c$ diminished by the radii of both spheres: $r = r_c - R_1 - R_2$.

Eq. (3). It is simply demonstrated that all its coefficients tend to $1/C_0$ for $\rho \to 0$, where $C_0$ is the total capacitance of the two conductors in electrical contact with each other (at $\rho = 0$). Therefore, the matrix determinant $D(\rho) = S_{11}(\rho) \, S_{22}(\rho) - S_{12}^2(\rho)$ tends to zero as well. Moreover, since $S_{ij}(\rho)$ are continuous and monotonic functions, their derivatives will tend to finite values. Let us consider, for the sake of simplicity, the first derivative of $C_{11} = S_{22}(\rho)/D(\rho)$, that is $C'_{11}(\rho) = S'_{22}(\rho)/D(\rho) - S_{22}(\rho)D'(\rho)/D^2(\rho)$, whose leading term, for $\rho \to 0$, is $-1/C_0 \, D'(\rho)|_0 \, \rho 2$. Similarly, it can be shown that the same holds for $C'_{22}(\rho)$ and $-C'_{12}(\rho)$. Hence their ratios must tend to one for small distance. It can be shown that the same holds for second derivatives as well. According to this result, our model of electrostatic interaction always reduces to Eq. (1) for small enough probe-sample distance.

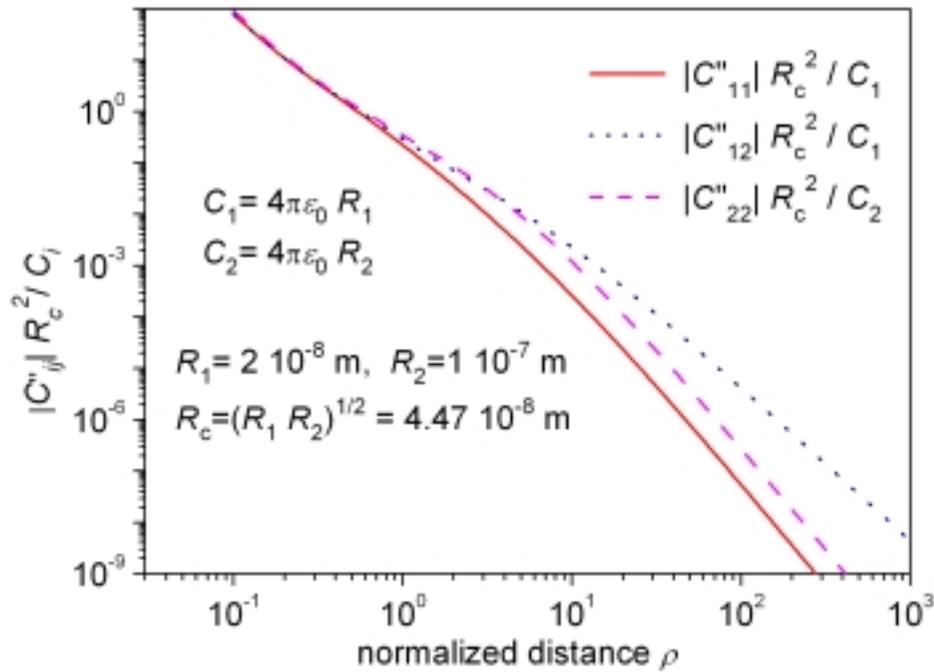

**Figure 2**: Simulated trend of $C''_{ij}$ as a function of normalized distance $\rho$, in the case $K_r = 5$.

We point out that EFM is usually operated by lifting the probe over the surface with an offset height, in the attempt to avoid topographic influence on the measurements [8]. This, however, often releases the condition of small probe/sample distance needed to validate Eq. (1), leading to the need of more accurate models to describe EFM findings. In this respect, as shown above, the limit of Eq. (1) – overcome by our model – is the inability to describe force gradient inversion at a given probe-sample distance.

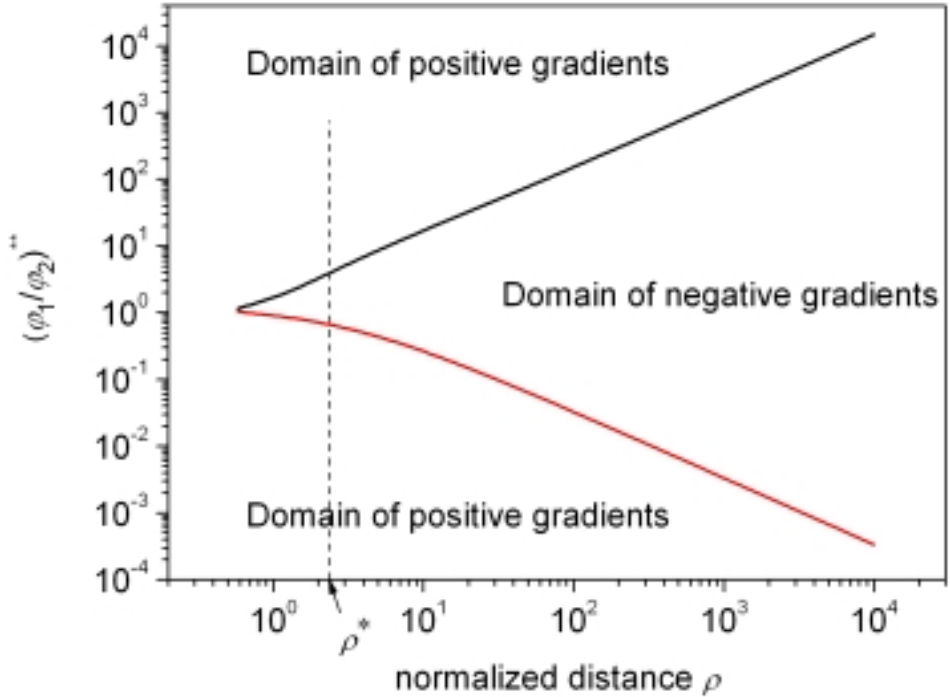

Figure 3: Simulated trend of $(\varphi_1/\varphi_2)^{**}$ values (where force gradient is null) as a function of $\rho$, for $K_r = 5$. The maximum inverted force gradient occurs at $\rho^*$ (dashed line).

Let us investigate the range of probe-sample distance where this phenomenon occurs, to assess its relevance to practical applications of EFM. In Fig. 3 we show, in the case of two spheres chosen as in Fig. 2, the trend with $\rho$ of $(\varphi_1/\varphi_2)^{**}$ solutions (Eq. (6)), that limit the regions of potential ratios $\varphi_1/\varphi_2$ where force gradients assume negative values. Such solutions are always positive, corresponding to potentials of same sign, as expected. The value of maximum force gradient $|F'_z(\varphi_1/\varphi_2)|_{max}$, is located where the derivative of $F'_z$ with respect to $\varphi_1/\varphi_2$ is null. Such maximum gradient, normalized to $\varphi_2^2$, is reported in Fig. 4 as a function of $\rho$. For size ratio $K_r$ significantly bigger than one we found that the distance at which the maximum gradient occurs is comparable to the radius $R_2$, that in the case of our simulations represents the sample size. The large distance necessary for inversion with high $K_r$ seems to be in agreement with the results of Ref. [5], where the sample is a wide and flat metal electrode, and noticeable effects of force gradient inversion are only found starting from a distance of about 1-2 µm.

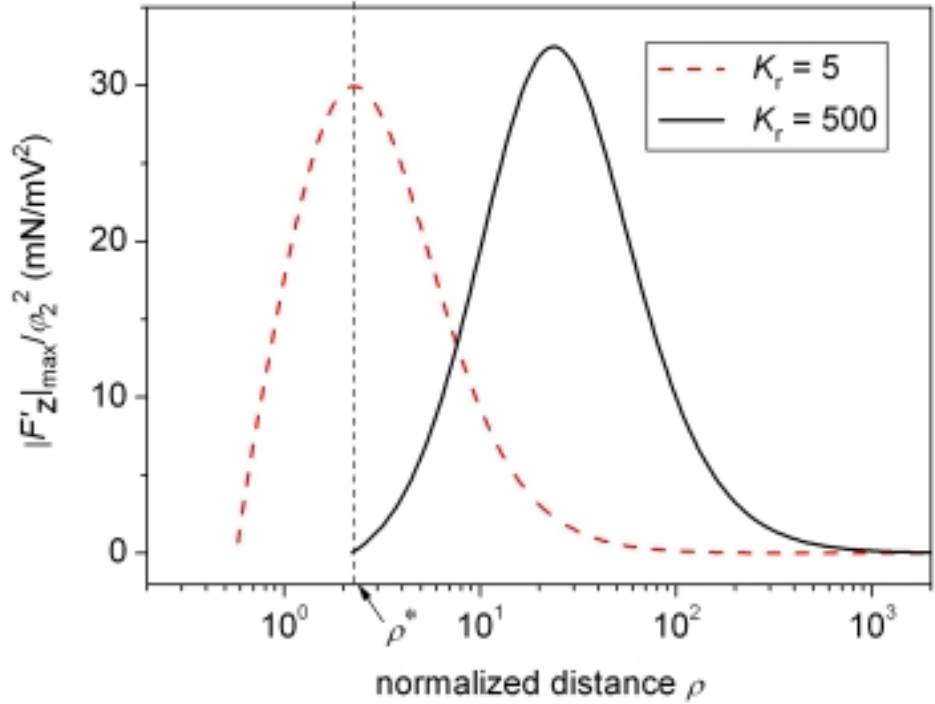

**Figure 4**: Simulated trend of $|F'_z(\varphi_1/\varphi_2)|_{max}$ vs. $\rho$ for two conducting spheres with radii $R_1 = 20$ nm and $R_2 = 100$ nm ($K_r = 5$) or 10 µm ($K_r = 500$).

All features demonstrated for the force gradient are also valid for the force, which also depends on potentials similarly to Eq. (5), by replacing $C''_{ij}$ with $C'_{ij}$. This will allow to extend our analysis to KPM measurements, related to first derivatives of capacitance coefficients. To better illustrate the investigated inversion phenomenon for both force and force gradient, Fig. 5 shows simulated values of such quantities as a function of $\varphi_1/\varphi_2$, in the case described in Fig. 3, at distance $\rho^*$ where the largest maximum force gradient is exhibited, to better evidence the effect. When potentials have opposite sign ($\varphi_1/\varphi_2 < 0$), force is always attractive (negative). When potentials have the same sign ($\varphi_1/\varphi_2 > 0$), force is repulsive in the region $0.6 < \varphi_1/\varphi_2 < 7$ only, and attractive outside such interval. Indeed, by applying high enough potential on one conductor, i.e. when $\varphi_1/\varphi_2$ approaches $0^+$ or infinity, force may become attractive even when potentials are of equal sign, because of the dominance of induced charges. The inversion interval for forces is larger than the one for force gradients, as also visible in the case of Fig. 5. Therefore, for a given probe-sample distance, there are more potential values leading to fall in the inversion region when operating a force-based technique, as KPM, than with EFM that is instead based on force gradients.

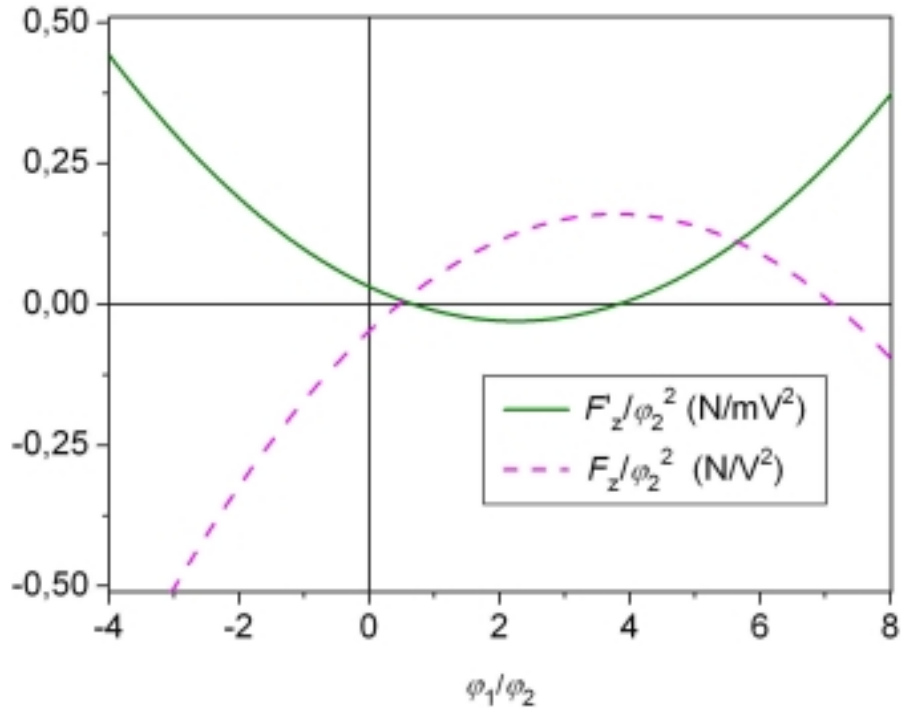

**Figure 5**: Simulation of normalized force ($F_z/\varphi_2^2$) and force gradient ($F'_z/\varphi_2^2$) vs. $\varphi_1/\varphi_2$ in the case of $R_1$ = 20 nm and $R_2$ = 100 nm, corresponding to $K_r = 5$, at distance $\rho^*$ as in Fig. 3.

At this point, we are also able to explain the ideal experiment proposed in the introduction. Indeed, we observe that, for the symmetric situation of Fig. 1a, $Q_1 = Q_2$ implies $\varphi_1/\varphi_2 = (C_{22} - C_{12})/(C_{11} - C_{21}) = 1$.

Since no generators are connected, the total charge on each conductor is conserved, and therefore $U_{tot} = U_{el}$. Furthermore, $C'_{11} < -C'_{12}$ as it can be evinced from both the derivative of the capacitance matrix in (8) at large distance and from simulations for small distance (not shown). Therefore, $F_z = -dU_{el}/dz = -(C'_{11} - C'_{12})\varphi_1^2 > 0$, that is, repulsive. In the case of Fig. 1b, potentials on the two spheres are about the same, and a repulsive force is present as above. In both such cases, repulsive force is due to the fact that $\varphi_1/\varphi_2 \sim 1$, and therefore the system is necessarily within the region of force inversion.[3] Instead, inversion would not be explainable at all according to Eq. (1), where the sign of force is only determined by the sign of the derivative of system capacitance $C'$, that by no means can be inverted by varying distance. When distance decreases, $\Delta' \to 0$ and inversion is not possible anymore. In such case, force is always attractive and Eq. (1) becomes valid. Therefore, our analysis can also provide limits of validity for Eq. (1). Indeed, it can be demonstrated that there always exist a distance where the force becomes zero while turning from attractive to repulsive; such distance $\rho_0$ is such that:

$$\frac{S'_{11}(\rho_0)}{S'_{12}(\rho_0)} = -\frac{2Q_1 Q_2}{(Q_1^2 + Q_2^2)}. \qquad (9)$$

To check predictions of our model as well as its practical relevance for common experiments, we have carried out EFM measurements by using two different sample geometries, that could be described in simulations by two different size ratios $K_r$. An AFM from Veeco (Mod. MultiMode Nanoscope IIIa with Extender Electronics Module) has been used for our tests. All measurements were performed with AFM probes suitable for EFM (Veeco, Mod. SCM-PIT7, *Pt-Ir* coated, cantilever spring constant $k \sim 2.8$ N/m, resonant frequency $v_0 \sim 70$ kHz, nominal tip curvature radius $\sim 20$-$25$ nm) in controlled humidity environment (<< 4% RH, the limit sensitivity of our hygrometer), to improve repeatability of measurements. EFM scans have been performed on gold stripe electrodes (5 μm wide, ~55 nm thick) deposited on a $SiO_2$ substrate, and separately on semiconducting nanowires (NW) of *InAs* (diameter ~67 nm, length ranging from 2 to 4 μm) deposited on them, as sketched in Fig. 6. Potentials $\varphi_1$ and $\varphi_2$ were applied to the probe and electrode, respectively. Charging experiments as well as KPM scans (not shown) indicated that our NWs behave as conductors, at least concerning quasi-static electric properties.

---

[3] It can be shown from Eq. (6) that the two solutions $(\varphi_1/\varphi_2)^{**}$ are reciprocal to each other, and that they both tend to one for $\rho \to 0$.

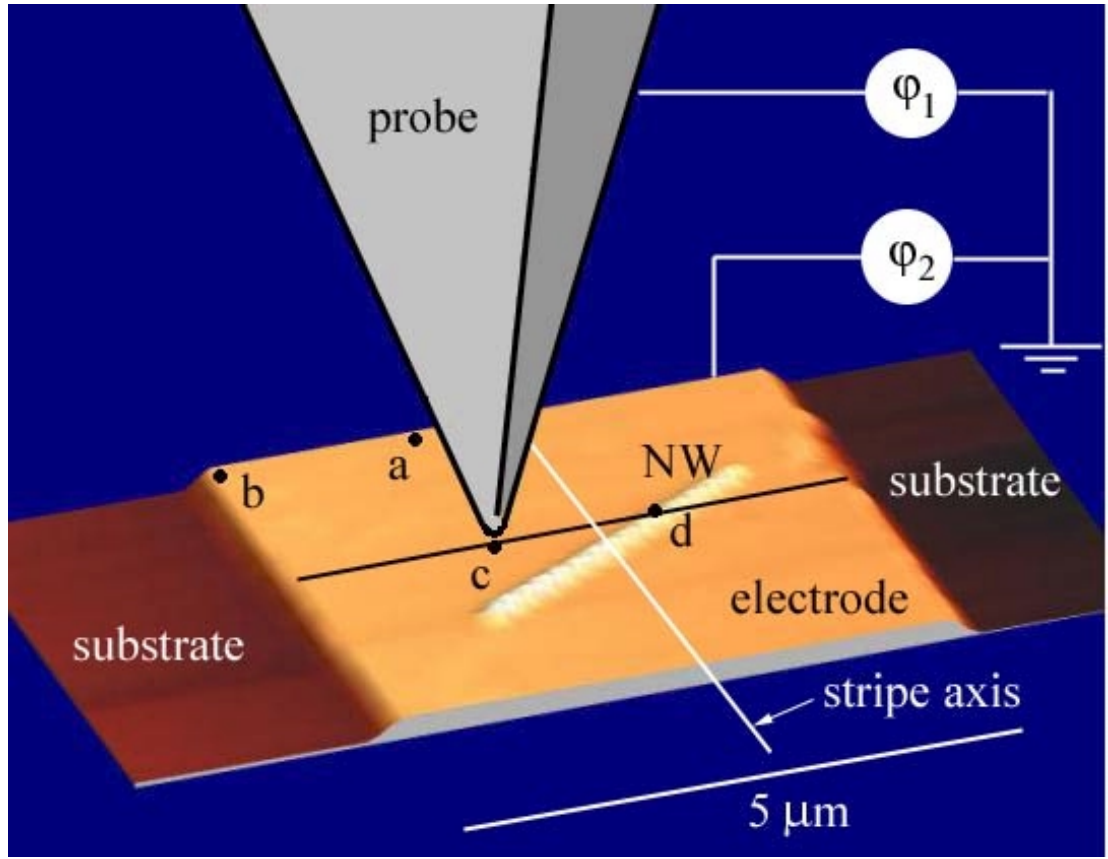

**Figure 6**: Experimental geometry. The AFM topographic image drawn in 3D [13] shows the NW deposited on the *Au* electrode, on top of the $SiO_2$ substrate. The *Pt-Ir* coated EFM probe is sketched in scale. Measurement points are indicated by the letters (a) to (d).

EFM measurements were performed in the Interleave "Linear mode" of the Nanoscope IIIa. Such mode consists in a first pass of each scan line in AFM TappingMode™ to measure a topographic profile, and a second pass along a linear trajectory – at a given height $h$ – parallel to the average of such profile. This mode has been chosen in order to minimize $z$-motion artifacts typically affecting probe microscopy techniques sensitive to long-ranged interactions [12]. Both probe and sample were grounded during the TappingMode pass, to minimize charge transfer effects. Bias was instead applied during the Interleave pass, when the EFM measurement is performed. We used probe oscillation amplitude $A \sim 10$ nm with typical lift height $h = 150$ nm, fulfilling fairly well the linearity condition for the force gradient. In Fig. 6, an actual tapping mode topographic image of the nanowire on the gold electrode is shown, while the probe is just sketched with the true proportions. The cantilever axis was nearly orthogonal to the stripe electrode axis.

Frequency shift values as a function of $\varphi_1$ and $\varphi_2$ were measured above the bare electrode as well as above the nanowire. Determination of coefficients $C''_{ij}$ has been conducted by a procedure aimed to eliminate influence of environmental measurement conditions, as follows. Data for $\varphi_2 = 0$ and varying $\varphi_1$ ($\varphi_1 = 0$ and varying $\varphi_2$) have been fitted to a second-order polynomial, whose quadratic coefficient corresponds to ½ $C''_{11}$ (½ $C''_{22}$). Data for a fixed $\varphi_2 \neq 0$ and varying $\varphi_1$ are then used to determine $C''_{12}$ by a linear fit of the function $(F'_z(\varphi_1,\varphi_2) - F'_z(\varphi_1,0) - F'_z(0, \varphi_2)) / \varphi_2$ where the linear fit coefficient corresponds to $C''_{12}$. It can be checked that this procedure allows the determination of capacitance coefficient derivatives by getting rid of the influence of work functions $V_i$ of each conductor, that act as to shift their potentials to $\varphi_i + V_i$.

| Position, height | $C''_{11}$ [μF/m²] | $C''_{22}$ [μF/m²] | $C''_{12}$ [μF/m²] | $\Delta''$ |
|---|---|---|---|---|
| (a), 95 nm | 56±2 | 54±1 | -57±3 | 0.1±0.1 |
| (b), 95 nm | 55±2 | 64±1 | -63±3 | 0.1±0.1 |
| (a), 150 nm | 59.2±0.6 | 53.1±0.8 | -57±3 | 0.1±0.1 |
| (c), 150 nm | 33.0±0.1 | 31.4±0.1 | -38±4 | 0.3±0.1 |
| (d), 150 nm | 42.6±0.2 | 43.1±0.1 | -52±2 | 0.3±0.1 |

**Table I**: Fit results for $C''_{ij}$ and $\Delta''$ at positions indicated in Fig. 6.

The obtained coefficients at various positions and heights are reported in Table I. Reproducibility of coefficient values between different sets of scans resulted around 5%. Therefore, our method allows, at least in principle, to determine $C''_{ij}$s regardless of the incipient work functions $V_i$, that might be influenced by ambient humidity and contaminants.

Fig. 7 shows the quadratic curves for $F'_z$ obtained by just inserting into Eq. (5) the capacitance coefficients of Table I. The analysis conducted above can be fully applied to such curves. Inversion phenomenon, corresponding to $\Delta'' > 0$, is evident especially for the cases close and above the NW (c) and (d), while the curve corresponding to the bare electrode (a), where $\Delta''$ is closer to zero, is almost tangent to the abscissa axis. Indeed, with reference to Table I, we see that on the bare electrode (a) $C''_{ij}$s are rather similar to each other, at both $h = 95$ and 150 nm. When the probe is placed close to the nanowire (c), $C''_{ij}$s tend to decrease and to become different to each other ($\Delta'' = 0.3$). Placing the probe exactly above the nanowire (d) such differentiation is consolidated, although coefficient values tend to increase again.[4]

We observe that for measurements on NWs there exists a range of $\varphi_1/\varphi_2$ values where the force gradient changes sign, that is the region of negative $F'_z$ in Fig. 7. Consequently, these measurements cannot be correctly described by Eq. (1). Indeed, due to the rather large probe-sample distance employed; $\Delta''$ is markedly greater than zero. We speculate that changes of $C''_{ij}$s by laterally displacing the probe are mainly due to the contribution of the full NW approaching the probe cone. The contribution of the underlying electrode provides smaller change, because of its higher translational symmetry.

We could also check that proximity of the electrode border gives much smaller deviation from the value at position (a) than the one due to the presence of the NW at position (c) and (d). The probe apex contribution plays a bigger role when its distance from the NW becomes comparable to the radius of the NW itself, by increasing $C''_{ij}$s again (d).

---

[4] We remark that $h$ indicates the distance between the probe and the electrode plane, therefore at position (d) the probe-NW distance will be smaller than $h$.

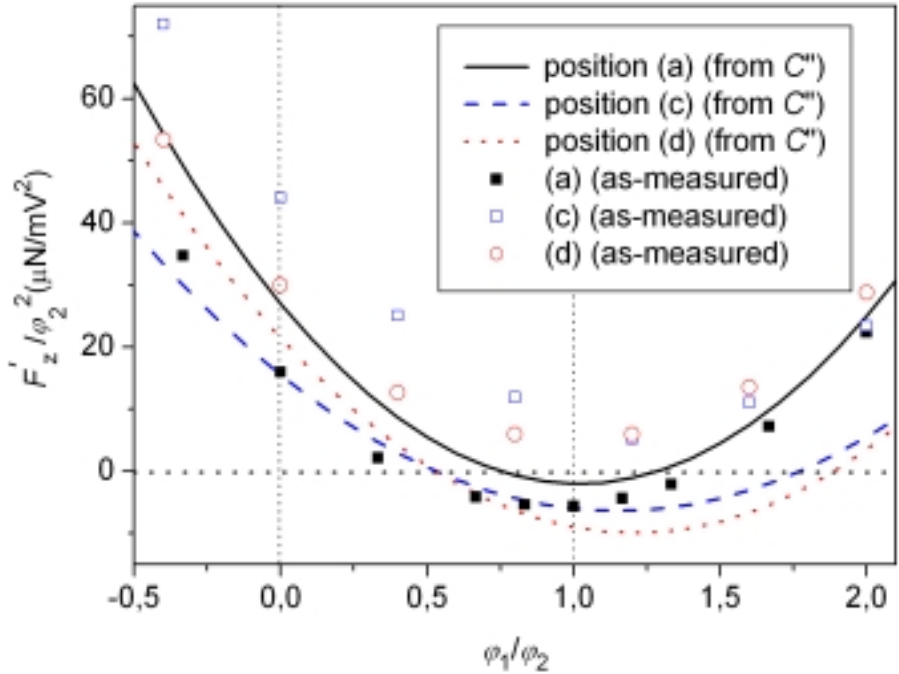

**Figure 7**: Quadratic $F'_z$ curves obtained by inserting the fitted $C''_{ij}$ of Table I into Eq. (5), plotted vs. $\varphi_1/\varphi_2$ at the positions indicated in Fig. 6 (solid lines). Some of the as-measured $F'_z$ values (dots), clearly presenting poor correspondence with such quadratic curves.

For comparison, the dots in Fig. 7 represent instead some of the as-measured force gradient values, from which capacitance coefficient derivatives have been calculated by fitting. We observe that such measurements have poor relation with the quadratic curves previously shown. This is expected, since the effect of work functions $V_i$ and of surface contamination provides deviation from the ideal curve described by Eq. (5), that has been derived regardless of the effect of work functions. Therefore, experimental results based on bare force gradient measurements will be subjected to deviations, by ambient conditions and humidity, more strongly than capacitance coefficients themselves. This is because the fitting procedure developed to obtain $C''_{ij}$s was tailored in such a way to get rid of the dependency on $V_i$, and actually, the obtained values of the capacitance coefficient derivatives are much more reproducible (about 5%) than the as-measured force gradients, as evident from Fig. 7. However, this issue is beyond the aim of the present work.

Let us now analyze some implications of our model for local potentiometry and KPM. As stated in the introduction, in KPM the value of DC potential $\varphi_1$ applied to the probe is regulated in order to null the electrostatic force component $F_{mod}$ oscillating at the modulation frequency $\nu_{mod}$ of the applied AC potential. By using our formalism, it is easily seen that the condition $F_{mod} = 0$, used in KPM, implies:

$$\varphi_1(z) = -\frac{\sum_{j=2}^{N} C'_{1j}(z)\varphi_j}{C'_{11}(z)} \tag{10}$$

We notice that Eq. (10) differs from the result presented in Ref. [10] in the denominator ($-C'_{11}$ instead of $\sum_{i=2}^{N} C'_{1i}$). As we have seen in the case of EFM for two conductors, such difference is not important at small distance (where $C'_{11} \to -C'_{12}$), but may become significant as soon as the microscope operates at such a distance for which $C'_{ij}$ differentiate. Moreover, Eq. (19) confirms that Kelvin potential $\varphi_1$ should be a function of distance, as already reported [10,14].

Finally, we observe that the condition $F_{mod} = 0$ can be also written as $\sum_{i=1}^{N} C'_{1i} \varphi_i = 0$, and by using Eq. (2) for the probe charge $Q_1$ one finds:

$$\frac{dQ_1}{dz} = \sum_{i=1}^{N} \frac{dC_{1i}}{dz}\varphi_i = 0 \tag{11}$$

the same condition used to measure Kelvin potential in macroscopic experiments, where modulated currents are nulled by properly adjusting DC bias between vibrating plates of a plane capacitor [3].

Therefore, nulling of $F_{mod}$ in KPM is equivalent to nulling currents at frequency $\nu_{mod}$ induced in the probe by the applied AC potential.

In conclusion, commonly used models for probe-sample electrostatic interactions in EFM and KPM experiments are not general enough to describe all possible experimental configurations, especially when probe-sample distance becomes comparable to the characteristic size of the structures under investigation. We propose a more general model in terms of capacitance coefficients in EFM and KPM that represents a more reliable basis for quantitative evaluations of local electric response at surfaces of realistic nanostructures. This includes sign inversion phenomena, not described by approximated expressions for electrostatic force and force gradient. In particular, our method can be used to obtain capacitance coefficient derivatives $C''_{ij}$ providing a description of the sample that is less influenced by environmental conditions than the bare as-measured force gradients. Our model also provides deeper insight of the physical meaning of local potentiometry measurements, by also yielding a more accurate expression for the local Kelvin potential in terms of applied potentials.

Financial supports from the INFM-CNR "SEED 2007" project, and from the "CIRV" project of University of Pisa are acknowledged. We also thank P.A. Rolla for useful discussion, S. Roddaro for supplying semiconductor nanowires, and M. Bianucci for technical support.